# About the Strain-Coupled Molecular Dynamics in the Ferroelastic Phase Transition of TMACd(N$_3$)$_3$

A. Nonato[1*], R. X. Silva[2], C.C. Santos[3], A. P. Ayala[4], C.W.A. Paschoal[4]

[1]Coordenação de Ciências Naturais Física, Universidade Federal do Maranhão, Campus do Bacabal, 65700-000, Bacabal - MA, Brazil

[2]Centro de Ciência e Tecnologia em Energia e Sustentabilidade, Universidade Federal do Recôncavo da Bahia, 44085-132, Feira de Santana-BA, Brazil

[3]Departamento de Física, Universidade Federal do Maranhão, Centro de Ciências de Exatas – CCET, 65080-805, São Luís - MA, Brazil

[4]Departamento de Física, Universidade Federal do Ceará, Campus do Pici, 65455-900, Fortaleza - CE, Brazil

[*]Corresponding author: Tel: +55 98 982054803

E-mail address: ariel.nonato@ufma.br (A. Nonato)

# ABSTRACT

Tetramethylammonium (TMA) cadmium azide, it is a new perovskite-like compound which undergoes a series of first-order phase transitions including a ferroelastic transition above room temperature. Understanding the order–disorder structural phase transition (SPT) mechanism in hybrid organic–inorganic perovskites (HOIPs) is crucial for designing new compounds with enhanced barocaloric efficiency, as well as unlocking other multifunctional properties. In this paper, we employed the energy fluctuation (EF) model to analytically investigated the linewidth of Raman modes in TMACd(N$_3$)$_3$ near to the critical phase transition temperature ($T_C = 322\ K$), aiming to gain insights into molecular dynamics around the SPT. The temperature dependence of the strain, used as an order parameter, was obtained using the appropriate thermodynamic potential for the first-order phase transition in TMACd(N$_3$)$_3$ in terms of Landau expansion, which can be successfully employed to model first-order ferroelastic phase transitions. We show that the EF model suitably captures the behavior of the Raman linewidths in the vicinity of the structural phase transition in TMACd(N$_3$)$_3$. The activation energies obtained for TMACd(N$_3$)$_3$ are comparable to those of DMACd(N$_3$)$_3$ as well as to $k_B T_C$. Additionally, the temperature dependence of the relaxation reveals that the torsional and librational modes require longer to renormalize after the phase transition in TMACd(N$_3$)$_3$ when compared with DMACd(N$_3$)$_3$. The discussion based on these new parameters through the EF model provides a new perspective for understanding molecular dynamics in systems undergoing order-disorder phase transitions, particularly in ferroelastic transitions, where order-disorder mechanisms are coupled to symmetry-breaking lattice distortions.

**Keywords:** Raman linewidths; Relaxation time; Hybrid perovskite; Raman spectroscopy; Ferroelastic phase transitions.

# 1. INTRODUCTION

The possibility of combining different metallic cations (A and B) and organic and inorganic anionic ligands (X) in perovskite-like structures, with the formula $ABX_3$, has been a path widely explored in the search for new and multifunctional materials [1]. Hybrid organic-inorganic perovskites (HOIPs) have attracted a lot of attention in recent decades and incredible advances have been made in their application as luminescent [2,3], photovoltaic [4], ferroelectric [5,6], and barocaloric [7,8] materials. The A-site of the perovskite is usually occupied by an organic amine cation, like methylammonium ($MA^+$), dimethylammonium ($DMA^+$), trimethylammonium ($TrMA^+$), tetramethylammonium ($TMA^+$), ethylammonium ($EA^+$), formamidinium ($FA^+$), among others. In general, the X ligands define the subgroups of HOIPs, which mainly include halogen anions, azides ($N_3^-$), formates ($HCOO^-$), cyanides ($CN^-$), and dicyanamides ($N(CN)_2^-$), forming the anionic crystal lattice whose larger cavity is occupied by the A cation, while the octahedral cavity is occupied by the B ion, a divalent transition metal ($M^{2+}$) and sometimes a bimetallic combination ($M^+$-$M^{3+}$) [9].

The insertion of organic moieties in the composition of perovskite structures, particularly the elongated X ligands, increases the degrees of freedom within the crystal lattice, enabling atypical distortions compared to purely inorganic compounds, such as unconventional octahedral tilts and columnar shifts [10,11]. In addition, HOIPs are relatively flexible, and it is known that the mobility of parts of the crystal structure under external stimuli can strongly influence their properties, such as polar cations whose displacements and reorientations are related to dielectric and ferroelectric properties [6], as well as the magnetic interactions established between transition metals and mediated by organic ligands [12,13]. These aspects make these compounds quite susceptible to structural phase transitions, which, in addition to octahedral tilts and shifts, also involve order-disorder processes, off-center displacements of A and B cations, and dynamic reorientations of the amine cations [14–20].

Among the mechanisms that drive structural phase transitions (STP) in HOIPs, order-disorder effects stand out for being quite common and often intense. Abrupt changes in structural ordering during an STP account for most of the entropy variation associated with the phase transition and are particularly desirable when dealing with barocaloric HOIPs [10,21]. HOIPs belonging to the azide subgroup, with the general formula $[(CH_3)_nNH_{4-n}][B(N_3)_3]$, where n=1-4, have gained a prominent position among barocaloric materials, as they present extensive first-order phase transitions, with large entropy variations that occur near room temperature [21–24]. However, the distinction between the mechanisms of phase transitions, the energies associated with each process, accommodation times, and temperature dependence of these processes is not trivial and requires multiple points of view and investigation techniques.

Particularly, TMACdN$_3$ has attracted much attention for displaying a singular above-room-temperature ferroelastic phase transition induced by the rotational dynamics of the $N_3^-$ bridges and the $[N(CH_3)_4]^+$ guests. Indeed, hybrid coordination polymers (CPs) benefit from their inherent structural tunability, enabling the design of new ferroelastic systems with multifunctional properties. Among the various types of coordination polymers (CPs), perovskite-like CPs stand out as particularly unique due to their structure, in which guest species are incorporated into well-matched host cages. These materials show great promise for the development of ferroelastic materials. However, reports on such CPs remain very scarce to date. [25–33]

In this sense, various studies have been conducted to explore the molecular dynamics and investigate the complex mechanisms of molecular reorientation that involve order-disorder processes in TMA and DMA-based compounds [23,24,34]. One approach to investigating molecular dynamics near the critical temperature (Tc) has been the application of molecular field theory. Several investigations have applied pseudospin-phonon coupling (PS) and energy fluctuation (EF) models to describe the damping behavior of Raman modes, as well as to calculate activation energies and relaxation times in systems undergoing order–disorder transitions [24-26]. The structural geometry and molecular dynamics of the organic molecules within the framework are important for determining the influence of temperature on the evolution of structural phase transitions as well as order-disorder effects in the organic-inorganic perovskite structure. Among the azide-based HOIPs, the compound $[N(CH_3)_4][Cd(N_3)_3]$ has shown potential barocaloric properties [24] and exhibits a ferroelastic phase transition which is driven by the swing of the azide bridge and rotational movements of the A cation [38]. However, to our knowledge, the molecular dynamics of TMACd(N3)3 have not been studied.

In this work, we applied the PS and EF models to investigate the temperature dependence of the lifetime of phonons in the lattice (FWHM) of the Raman modes of the organic-inorganic hybrid perovskite TMACd(N3)3, which undergoes two structural transitions ($\alpha \rightarrow \gamma \rightarrow \delta$) near room temperature, where the $\gamma\ (Pnma) \rightarrow \delta(Pm3-m)$ transition (from orthorhombic to cubic symmetry) is driven by significant structural modifications [38,39]. This phase transition is presumed to be ferroelastic, with an "Aizu-type" $m3mFmmm(pp)$, [40,41] where the number of equivalent unique ferroelastic directions is up to 12. The letters "$pp$" in parentheses indicate that the two dyad axes of the ferroelastic crystal are aligned with two of the tetrad axes of the prototype structure. From the order parameter derived using the standard first-order solution of a Landau 2-4-6 potential, we obtained relaxation times and activation energies for several molecular vibrational modes near the critical temperature. Considering that TMACd(N3)3 undergoes a ferroelastic phase transition, we further explore how molecular dynamics couple with symmetry-breaking strain, introducing new parameters that offer deeper insight into the order–disorder mechanisms governing this transformation.

## 2. MODEL

To explain the orientational order transitions of $NH_4^+$ ions in the $NH_4Br$ crystal, Yamada *et al.* [42] proposed the Ising pseudospin-phonon coupled model. This model is more general than the simple pseudospin model as it considers the interaction between a pseudospin and a phonon, and can be used to obtain the temperature dependence of phonons in molecular crystals. The total Hamiltonian, considering the spin-phonon coupling term, is given by:

$$H = \frac{1}{2}\sum_{k,s}\left(p_{\vec{k}s}p_{\vec{k},s}^* + \omega_{\vec{k}s}^2 q_{\vec{k}s}q_{\vec{k}s}^*\right) - \frac{1}{2}\sum_{i,j}J_{ij}\sigma_i\sigma_j - \sum_{k,s}\sum_i \frac{\omega_{\vec{k}s}}{\sqrt{N}}g_{\vec{k}s}q_{\vec{k}s}\sigma_i e^{i\vec{k}\cdot\vec{r}_i} \quad (1)$$

In this equation, $q_{\vec{k}s}$ denotes the phonon coordinate for mode $s$ with $\vec{k}$ wave vector, $\omega_{\vec{k}s}$ is characteristic frequency and $p_{\vec{k}s}$ is the phonon moment. Thus, the first term gives the phonon energy, and the second term corresponds to the interaction energy between the nearest-neighbour spins with the spin variables $\sigma_i$ and $\sigma_j$ separated by $\vec{r}_{ij}$. The last term corresponds to spin-phonon interaction with coupling constant $g_{\vec{k}s}$. Finally, $N$ represents the number of unit cells in the molecular crystal.

Later, Matsushita [43] extended the model proposed by Yamada *et al.* to include more than one interaction of the phonon with the pseudospin. From these modifications in the microscopic Hamiltonian, Matsushita derived new expressions for the frequency and width of the phonons, being able to model the temperature dependence of the optical phonons on Ammonium Halides (CH$_4$Br and NH$_4$Cl). The expression for the linewidth of modes, according to Matsushita, shows a dependence on the dynamic scattering function of the pseudospins due to the pseudospin-phonon interaction with the $\nu_{th}$ phonon with wave vector $\vec{k}$, according to the equation:

$$\Gamma_{SP}(\vec{k}\nu,\omega) = \sum_{\vec{q}\nu'} \frac{|K_1(\vec{k},\vec{q},\nu,\nu')|^2}{8\omega\omega_0(\vec{k}-\vec{q},\nu')}$$

$$\times \left\{ \left[ \frac{n\left(\omega_0(\vec{k}-\vec{q},\nu')\right)}{n\left(\omega - \omega_0(\vec{k}-\vec{q},\nu')\right) + 1} + 1 \right] \times S(\vec{q}, \omega - \omega_0(\vec{k}-\vec{q},\nu')) \right.$$

$$\left. + \left[ \frac{n\left(\omega_0(\vec{k}-\vec{q},\nu')\right) + 1}{n\left(\omega + \omega_0(\vec{k}-\vec{q},\nu')\right) + 1} + 1 \right] \times S(\vec{q}, \omega + \omega_0(\vec{k}-\vec{q},\nu')) \right\} \quad (2)$$

where the term $K_1(\vec{k}, \vec{q}, \nu, \nu')$ represents an effective coupling constant of the $(\vec{k}, \nu)$ and $(\vec{k} - \vec{q}, \nu')$ phonons and a pseudospin with wave vector $\vec{q}$. $S(\vec{q}, \omega)$ is the dynamic scattering function of the pseudospins. The temperature dependence of the phonon linewidth for a phonon of frequency $\omega_0(\vec{k}\nu)$ is thus determined by the critical behaviour of the scattering function $S(\vec{q}, \omega \cong 0)$, as indicated by Matsushita [44]. Consequently, an anomalous behaviour of the scattering function in the vicinity of $T_c$ gives rise to an anomalous behaviour in the linewidth of the phonons. Laulicht and Luknar [45] investigated the temperature dependence of phonon linewidths near $T_c$ by introducing some approximations in Eq. (2), to simplify it. Based on these approximations, they derived an expression for the temperature-dependent phonon linewidth in the vicinity of the critical temperature, given by:

$$\Gamma_{SP} = \Gamma_0 + A\,(1 - Q^2)\ln\left[\frac{T_C}{T - T_C(1 - Q^2)}\right] \qquad (3)$$

In this equation, $Q$ corresponds to an order parameter (strain for ferroelastics), $\Gamma_0$ is the background linewidth, and $A$ is a fitted constant.

A different approach to calculate the temperature dependence of the linewidth $\Gamma_{SP}$ due to the pseudospin-phonon coupling was formulated to explain the critical broadening of the linewidth of the modes in Triglycine Sulphate (TGS) and Triglycine Selenate (TGSe) [46]. The observed critical broadening has been understood as due to the energy fluctuations (EF) of these modes [47,48]. Meanwhile, the observed shift of the phonon modes has been interpreted as due to changes in some spring constants in a molecular group when flipping motion between two equilibrium positions occur. Accordingly, the temperature dependence of the linewidth $\Gamma_{SP}$ for the EF model was defined by Schaack and Winterfelt [49], and subsequently presented by Laulicht as [45] follows:

$$\Gamma_{EF} = \Gamma_0' + A'\left[\frac{T(T - Q^2)}{T - T_C(1 - Q^2)}\right]^{1/2} \qquad (4)$$

where $\Gamma_0'$ is the background linewidth and $A'$ is a fitted constant, as defined earlier. The temperature dependence of the order parameter was derived by Matsushita [43] from the molecular field theory as:

$$Q = \left[3\left(1 - \frac{T}{T_C}\right)\right]^{\frac{1}{2}}, \; 0 < (T_C - T) < T \qquad (5a)$$

$$= 1 - 2\exp(-2T_C/T), \; T \ll T_C \qquad (5b)$$

$$= 0, \qquad\qquad\qquad T > T_C \qquad (5c)$$

where $T_C$ is the critical transition temperature. The expressions of the phonon frequency, linewidth Eqs. (3) and (4), and the order parameter (Eq. 5) has been widely used in previous studies [35,50–52]. Unfortunately, this expression is not appropriate for describing the order parameter in first-order transitions, as such transitions involve discontinuous changes that are not captured by continuous models based on critical exponents-based Landau theory.

In such cases, spontaneous lattice distortions provide a more reliable framework for analyzing structural behavior, especially in ferroelastic transitions involving symmetry breaking. These distortions can be quantified through the spontaneous strain tensor, which reflects variations in lattice geometry and may be related to the driving mechanism of the transition. To investigate the nature of the first-order ferroelastic cubic-to-orthorhombic transition, we calculated the spontaneous strain components relative to the extrapolated cubic parameter $a_0$. The non-zero components are defined as $e_1 = (a - a_0)/a_0$, $e_2 = (b - a_0)/a_0$, and $e_3 = (c - a_0)/a_0$, where $a$, $b$, and $c$ are the orthorhombic lattice parameters, and $a_0$ is the cubic lattice parameter extrapolated into the orthorhombic stability range (see Fig. 1a) [53]. Strains calculated from the lattice parameter data in this way are shown in Fig. S1. Values of $e_1$, $e_2$ and $e_3$ are in the range of ~-2.8 to +3.3% and giving a total volume strain ($\approx e_1 + e_2 + e_3$) of ~3.5%.

As highlighted by Zhao *et al.* [54], the order parameter associated with the cubic-to-orthorhombic symmetry change comprises multiple components, and therefore, a simple linear relationship between them is no longer expected. In phase transitions where the high-symmetry parent phase is cubic and the driving order parameter transforms according to a doubly degenerate representation ($E$), it is convenient to express the spontaneous strain in terms of symmetry-adapted components. This approach reduces the number of independent strain variables and establishes a direct link to the symmetry-breaking mechanism. In this framework, the spontaneous lattice distortion is characterized by two degenerate symmetry-breaking strain components, $e_t$ and $e_0$, adapted to the doubly degenerate irreducible representation $E$ [55]:

$$e_0 = e_1 - e_2 \tag{6}$$

$$e_t = \frac{1}{\sqrt{3}}(2e_3 - e_1 - e_2) \tag{7}$$

The factor $\frac{1}{\sqrt{3}}$ is included in the definition of et to ensure that the two strains are on the same scale. The adapted symmetry-breaking strain ($e_0^2 + e_t^2$) can be used as an order parameter to describe proper-ferroelastic phase transition. This adapted symmetry-breaking strain combination was successfully employed to model the order parameter associated with the first-order orthorhombic-to-cubic phase transition (P2$_1$3 → P2$_1$2$_1$2$_1$) in K$_2$Cd$_2$(SO4)$_3$ [55,56]. Accordingly, Fig. 1b captures the temperature dependence of the symmetry-breaking order parameter across the γ → δ transition in TMACd(N₃)₃, evidencing a discontinuity at $T_C$ consistent with a first-order ferroelastic phase transition. The order parameter used in Eq. (4) is considered in its normalized form, $Q/Q_{máx}$, to facilitate comparison between experimental and theoretical results. Also, $(e_0^2 + e_t^2)$ scales with $Q^2$, where Q is the driving order parameter and its variation can therefore be modeled by the standard first-order solution for a 246 Landau potential (See **Note 1** in the supplementary material) [55,57]:

$$(e_0^2 + e_t^2) = \frac{2}{3}(e_{0,0}^2 + e_{t,0}^2)\left\{1 + \left[1 - \frac{3}{4}\left(\frac{T-T_c^*}{T_{tr}-T_c^*}\right)\right]^{1/2}\right\} \qquad (8)$$

Here, $e_{0,0}$ and $e_{t,0}$ are the values of the order-parameter components in the orthorhombic phase at the equilibrium transition temperature, $T_{tr}$ is the transition temperature, $T_c^*$ the critical temperature and $T_c^*$ is the value of $T_c$ renormalised by coupling of the symmetry-breaking strains with $e_0$ and $e_t$. The least-squares fit shown in Fig. 2b gave the difference $T_{tr} - T_c^*$ is only 8 K (for $T_{tr}$ taken as 318 K), which is consistent with the behavior expected for weakly first-order transitions near a tricritical point, as predicted by the 246 Landau potential.

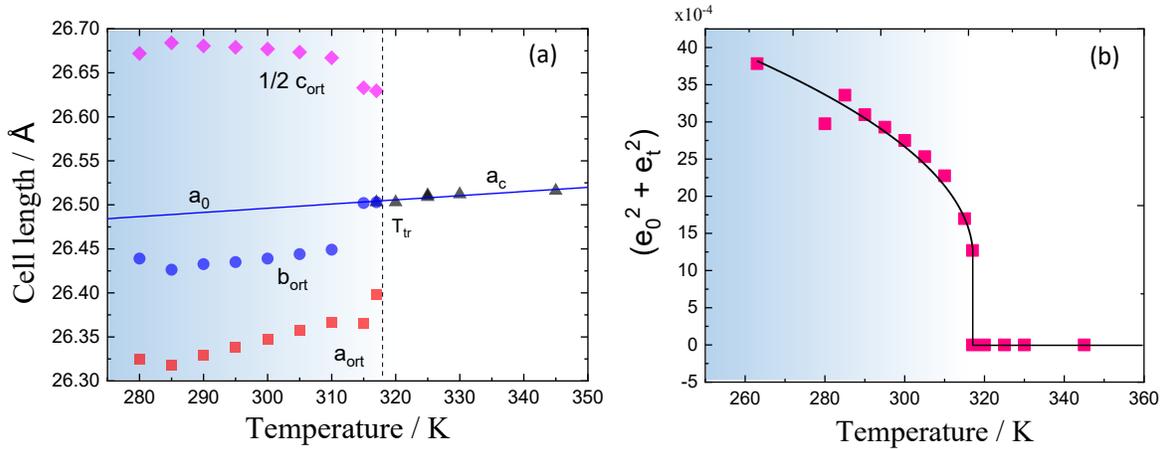

**Figure 1**. (a) Temperature dependence of the lattice parameters in TMACd(N$_3$)$_3$, extracted from ref. [39]. (b) Fits of $(e_0^2 + e_t^2)$ against temperature for the first-order transition in TMACd(N$_3$)$_3$. The curve represents solutions of a Landau potential with a negative quartic coefficient (b$^*$ < 0), as discussed in ref. [55]. The solid black line represents the best fit curve of the data below the ferroelastic transition point, modeled using Eq. (8), which describes the temperature dependence of the order parameter based on a first-order Landau potential.

Once we have the pseudospin-phonon coupled (Eq. (3)) and the energy fluctuation (Eq. (4)) models, we predicted the activation energy and the relaxation time as a function of temperature. Activation energy U can be determined as a function of temperature from the linewidth $\Gamma_{SP}$ using both models studied here according to the relation [58,59]:

$$\Gamma \cong \Gamma_{vib} + C \exp\left(\frac{-U}{k_B T}\right) \qquad (9)$$

where $\Gamma_{vib}$ is the vibrational damping (it is approximately zero as the transition temperature get closer), $C$ is a constant e $k_B$ is the Boltzmann constant. Assuming vibrational relaxation is almost zero near $T_C$ ($\Gamma_{vib} = 0$), Eq. (9) simplifies to an Arrhenius-like behavior, allowing activation energy U can be calculated according to:

$$\ln \Gamma \cong \ln C - \frac{U}{k_B} \cdot \left(\frac{1}{T}\right) \qquad (10)$$

where the slope of $\ln \Gamma$ with the $1/T$ (straight line) gives the activation energy with $\ln C$ as the intercept.

From the temperature dependence of the normalized order parameter $Q$ obtained from Fig. 1 and the linewidth, we can obtain the lattice relaxation time (τ) as a function of temperature for phonon modes as given below:

$$\tau = \frac{\Gamma}{Q^2} \qquad (11)$$

## 3. RESULTS AND DISCUSSION

The experimental linewidth data for selected vibrational modes presented in this work were extracted from our previous study, where we performed temperature-dependent Raman spectroscopy on the barocaloric hybrid perovskite [(CH₃)₄N][Cd(N₃)₃] [24]. Room temperature Raman spectra and mode assignments were previously reported in detail in Ref. [24], where experimental conditions, spectral analysis, and group theoretical assignments were fully discussed. In the present work, we reanalyze the linewidth data to explore critical broadening phenomena and their connection to strain-coupled molecular dynamics during the ferroelastic γ → δ phase transition in TMACd(N₃)₃. The linewidth

(FWHM) of the modes observed at 221 cm$^{-1}$ (LN$_3^-$), 275 cm$^{-1}$ (τCH$_3$), 751 cm$^{-1}$ (ν$_s$NC$_4$), 1360 cm$^{-1}$ (ν$_s$(ν$_1$)N$_3$), 2954 cm$^{-1}$ (ν$_s$CH$_3$) and 3032 cm$^{-1}$ (ν$_{as}$CH$_3$ ) (Fig. 2a-2f and Fig. S1 1a-1f)) for TMACd(N$_3$)$_3$ was analyzed as a function of temperature using the symmetry-adapted strain formalism and the first-order Landau 2–4–6 potential (Eq. 8). The temperature evolution of the linewidths was modeled using the pseudospin–phonon coupling (PS) model (Eq. 3) and the energy fluctuation (EF) model (Eq. 4), by fitting data for $T < T_C$ (see Table 1). In this work we chose to select the most intense modes from different regions of the spectrum observed for the TMACd(N$_3$)$_3$. The adjustments made were reasonable and agreed with the experimental data of the selected modes in the IT (intermediate phase - γ) close to $T_C$ (Fig. 2a-2f). Here, based on the analysis of the fitted parameters, we found that the EF model provided a better fit to the linewidth of L(N$_3^-$), τ(CH$_3$), ν$_s$(NC$_4$), ν$_s$(N$_3$), ν$_s$(CH$_3$), and ν$_{as}$(CH$_3$) modes ($T < T_C$) than the PS model (Fig. S2 1a-1f). The model deviations observed for temperatures below ~280 K are related to the structural phase transition ($\alpha \rightarrow \gamma$) not investigated here [60]. It should be highlighted that the energy fluctuation model is particularly suitable for describing systems exhibiting critical slowing down, where the phonon lifetime shortens and the linewidth (FWHM) increases significantly near the transition temperature due to enhanced order parameter fluctuations, even in the absence of noticeable frequency shifts. This result is in good agreement with previous Raman studies on TMACd(N$_3$)$_3$, which reported only minor frequency shifts for the symmetric stretching modes of both the methyl group (CH$_3$) and the azide anion (N$_3^-$) across the γ → δ transition. As previously discussed, the observed anomalies in the linewidth of the modes are due to the critical behavior of the scattering function $S(\vec{q}, \omega \cong 0)$ near $T_C$. The EF model successfully captures the critical broadening behavior of Raman modes near the structural transition. The abrupt increase in linewidth near $T_C$ is a hallmark of critical fluctuations associated with the order-disorder dynamics of the system, and the subsequent saturation reflects the stabilization of a high-symmetry phase. This highlights the sensitivity of the phonon lifetime to microscopic fluctuations, validating the EF framework for this ferroelastic transition. Therefore, in this work, we focus on the results obtained through solely the energy fluctuation model (EF). A comparison with PS model is given in the supplementary material.

The discontinuities observed in the linewidths of the modes (Fig. 2b-2f) mark the first-order structural phase transition in TMACd(N$_3$)$_3$ that occurred at $T_C$ = 322 K [24].

Interestingly, the bandwidth of the mode observed at 221 cm$^{-1}$ showed no discontinuity at the transition temperature $\gamma \rightarrow \delta$, which may be explained by the limited angular rearrangement of the N$_3^-$ group (N–N–N) during the $\gamma \rightarrow \delta$ phase transition, as this mode is predominantly associated with its bending vibration. In contrast, we observe that there is an abruptly increase in the linewidth of this mode near $T_C$ (Fig. 2a). Recently, we have shown that the discontinuous broadening observed for some modes in the transition $\gamma \rightarrow \delta$ can be attributed to short-range order-disorder effects and are sensitive to the vibrations of the $\tau(CH_3)$ $\nu_s(NC_4)$, $\nu_s(N_3)$, $\nu_s(CH_3)$ and $\nu_{as}(CH_3)$ groups [60]. Moreover, the weakening of hydrogen bonds between azide groups and the TMA cation could be the reason for the broad bands observed above that critical temperature [60]. Furthermore, structural transformations observed in $\gamma \rightarrow \delta$ phase transition have been assigned to the sway of the rod-like N$_3^-$ bridges as well as the rotation of the tetrahedron-like [N(CH$_3$)$_4$]$^+$ guest at different temperatures in TMACd(N$_3$)$_3$ [38].

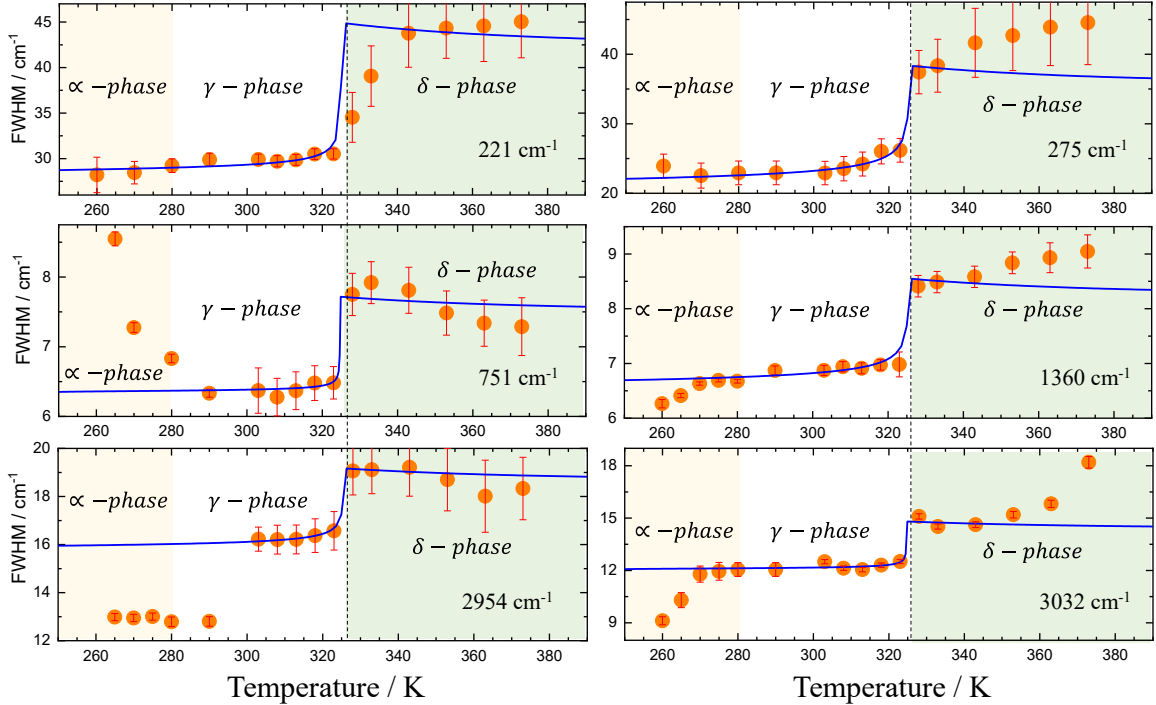

**Figure 2**. Temperature dependence of the full width at half maximum (FWHM) for selected vibrational modes of TMACd(N$_3$)$_3$. Solid blue lines represent Eq. (4) fitted to experimental data [24] according to the energy fluctuation (EF) model. The shaded regions indicate the structural phases: α-phase (beige), γ-phase (white), and δ-phase (green). The vertical dashed line at $T_C$ = 322 K denotes the ferroelastic $\gamma \rightarrow \delta$ phase transition, evidenced by an abrupt broadening of the linewidth. Here, we consider that above $T_C$, the order parameter $Q = 0$, as expected for the paraelastic phase. A second anomaly, observed below ~280 K, is associated with the α → γ transition.

**Table 1.** Fitting parameters values of the linewidth (FWHM) and the amplitude according to energy-fluctuation (EF) model (Eq.4) within the temperature intervals for TMACd(N$_3$)$_3$.

| Modes | | EF model | | |
|---|---|---|---|---|
| Assignment | w (cm$^{-1}$) | $\Gamma_0'$ | $A_0'$ | Temp. Interval (K) |
| L(N$_3^-$) | 221 | 28.45 ± 0.66 | 0.457 ± 0.025 | 260.4<T<343 |
| τ(CH$_3$) | 275 | 20.97 ± 0.61 | 0.483 ± 0.022 | 260.4<T<343 |
| ν$_s$(NC$_4$) | 751 | 6.35 ± 0.09 | 0.038 ± 0.003 | 290.0<T<343 |
| ν$_s$N$_3$ | 1360 | 6.57 ± 0.20 | 0.055 ± 0.007 | 269.8<T<343 |
| ν$_s$(CH$_3$) | 2954 | 15.82 ± 0.11 | 0.093 ± 0.003 | 270.2<T<343 |
| ν$_{as}$(CH$_3$) | 3032 | 12.07 ± 0.21 | 0.075 ± 0.007 | 269.8<T<343 |

The first result obtained from the application of the EF model is the estimation of relaxation times ($\tau$) associated with the renormalization of vibrational modes, which provide insight into structural phase transitions and molecular reorientational dynamics. This parameter can provide insights into the reorientation processes of molecules that can be associated with different physical properties. Thus, several previous works have focused on understanding related aspects the dynamics of the structural phase transition employing the spin-lattice relaxation time measurements [37,51,61]. Here, the relaxation times ($\tau$) of the L(N$_3^-$), τ(CH$_3$), ν$_s$(NC$_4$), ν$_s$ (N$_3$), ν$_s$(CH$_3$) and ν$_{as}$(CH$_3$) modes were calculated by applying Eq. (11) as a function of temperature below $T_C$ by using the calculated values of the linewidth (FWHM) from Eq. (3) and (4) and the order parameter obtained from Fig. 1b, as plotted in Fig.3 (3a-3f). The calculated values of relaxation time for all selected modes in TMACd(N$_3$)$_3$ abruptly increase with temperature up to 322 K in the γ phase, being lower and relatively constant far from $T_C$, as expected for a dynamically stabilized γ-phase where molecular disorder is reduced [62]. We found that the vibrational modes L(N$_3^-$) and τ(CH$_3$) have longer relaxation times, about 10.2 ns and 7.7 ns, respectively, as determined from our calculations. Here a longer relaxation time indicates that these modes take a longer time to renormalize phonon energy in the molecular reorientational motion. It is well established that low-frequency torsional modes, such as the methyl group twisting τ(CH$_3$), and internal librations involving heavier moieties like L(N$_3^-$), typically exhibit longer relaxation times compared to high-wavenumber stretching modes. This behavior has been attributed to the weaker vibrational coupling and the lower density of accessible final states for energy dissipation.

For instance, studies on methanol have shown that C–H stretching modes relax on sub-picosecond timescales, while torsional motions persist over several picoseconds due to their inefficient energy transfer pathways [63,64]. These findings support the experimental observations in this work, where τ(CH$_3$) and L(N$_3^-$) modes display slower dynamics, consistent with their vibrational nature. Furthermore, the increase in relaxation time may reflect the resistance of the system to accommodate strain fluctuations, highlighting the coupling between lattice distortions and molecular dynamics [62].

On the other hand, a shorter relaxation time, as observed for the modes ν$_s$(NC$_4$)(~2.2 ns) and ν$_s$(N$_3$) (~2.3 ns), indicates these phonons are renormalized faster after the phase transition. The results herein are comparable to those calculated for DMACd(N$_3$)$_3$ [35]. However, the relaxation times obtained in the present study are noticeably higher than those reported by Kurt at about 174 K for the modes τ(CH$_3$) (~1.2 ns), ν$_s$(NC$_4$) (~0.28 ns) and ν$_s$(N$_3$) (~0.35 ns) [35]. This difference may be attributed to the larger size of the TMA molecule compared to DMA, which can hinder molecular reorientational motion and enhance the coupling with the lattice strain, leading to slower relaxation dynamics. Additionally, the lower measurement temperature in Kurt's study for DMACdN$_3$ may also contribute to the shorter relaxation times observed.

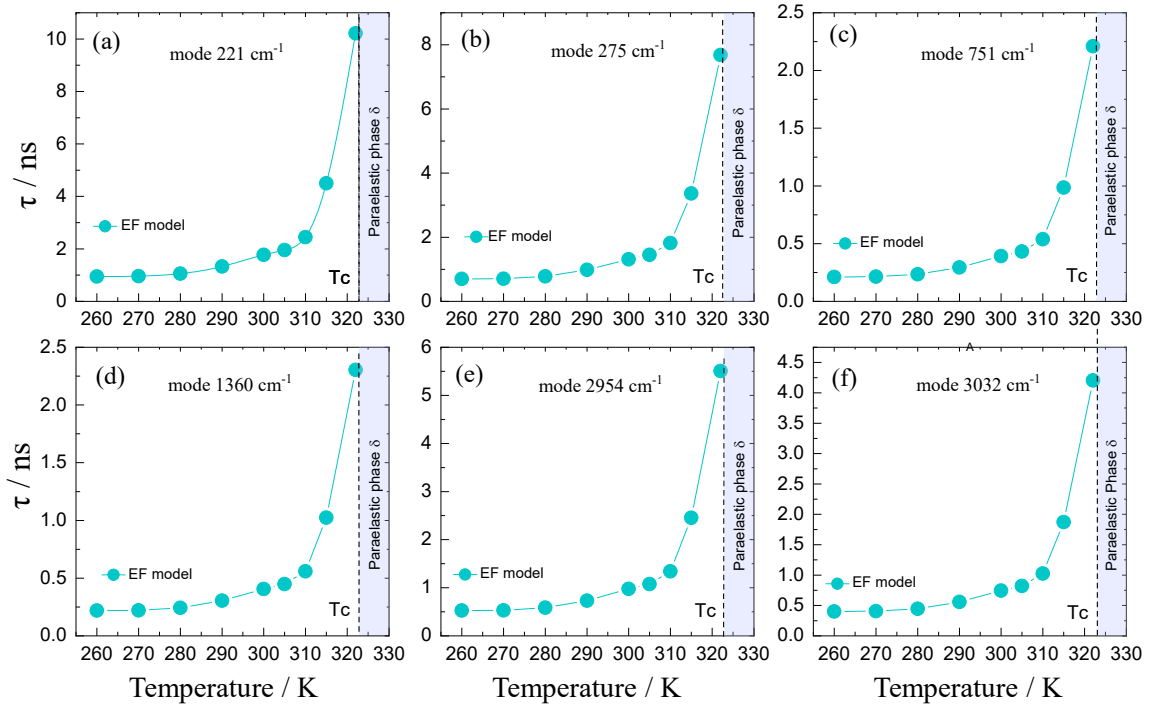

**Figure 3**. Relaxation time calculated by Eq. (11) for EF model of (a) L(N$_3^-$), (b) τ(CH$_3$), (c) vs(NC$_4$), (d) ν$_s$(N$_3$), (e) ν$_s$(CH$_3$) and (f) ν$_s$ (CH$_3$) modes as a function of temperature for TMACd(N$_3$)$_3$.

Using the linewidth (Γ) values calculated via the EF model, we extracted the values of the activation energy U using Eq. (10). Activation energy represents the potential barrier that separates a molecular group between two possible orientational states as a function of that state's energy and the temperature of the system. Thus, higher activation energy compared to the $k_BT$ term, in statistical terms of the Boltzman distribution function, represents a lower probability of molecules to reorient to the new state [65]. On the other hand, when the energy activation is lower than $k_BT$, molecular reorientation becomes more probable. The activation energy values obtained through the EF model are listed in table 2. These results are in good agreement with those observed for DMACd(N$_3$)$_3$ obtained through molecular field theory [35]. Activation energies for all Raman modes in TMACd(N$_3$)$_3$ increased as the temperature range approach to $T_c$. The lowest activation energies were found for the modes ν$_s$(NC$_4$) (~ 5.9 meV), ν$_s$(CH$_3$)(~ 3.2 meV) and ν$_{as}$(CH$_3$) (~ 2.8 meV). The last two values are approximately the same, indicating that the symmetric and asymmetric stretching modes of the CH$_3$ group are thermally activated in a similar way and exhibit activation energies that are practically temperature independent. Otherwise, the activation energies calculated for the τ(CH$_3$) mode (~24.7 meV) and the ν$_s$(N$_3^-$) mode (~17.4 meV) are relatively close to the thermal energy at the transition temperature, $k_BT_C$ (28 meV). This comparison is important because it indicates that the molecular motions associated with these modes are thermally accessible near the phase transition and may actively contribute to the ferroelastic behavior by facilitating local structural rearrangements or coupling with lattice strain fields that drive the symmetry-breaking process.

Therefore, we note that the activation energy calculated for the τ(CH$_3$) mode (~24.7 meV) in the TMACd(N$_3$)$_3$ is approximately three times higher than that calculated for the DMACd(N$_3$)$_3$ (~7.28 meV). Furthermore, the activation energy calculated for the symmetric stretching mode ν$_s$(N$_3$) (~8.8 meV) and ν$_s$(NC$_4$) (~5.9 meV) in TMACd(N$_3$)$_3$ are higher than those observed for DMACd(N$_3$)$_3$. This reveals τ(CH$_3$), ν$_s$(NC$_4$) and ν$_s$(N$_3$) modes in TMACd(N$_3$)$_3$ have a higher potential barrier compared to those in DMACd(N$_3$)$_3$. Once the driving force of the γ→ δ phase transition in TMACd(N$_3$)$_3$ is not clear yet [25], and believing that a deep understanding of the observed differences in the activation energies of these systems can provide insights into these contrasts, we have highlighted some of the most prominent differences between these structures. The first one is related to the size of the cations that occupy the cavity in these azide structures. Usually, the stability of these structures is defined through the tolerance factor ($t$) that involves the effective ionic radius of these molecules. The dimethylammoium molecule has an effective ionic radius of 272 pm ($t = 0.919$), a polar molecular structure, and tends to present more dynamic behavior and an anisotropic orientation relative to the framework [66], whereas the tetramethylammonium molecule is larger, with an effective ionic radius of 292 pm ($t = 0.963$), a tolerance factor near to that predicted for an ideal perovskite ($t = 1.0$), also, TMA has a larger spheroidal molecular structure. This is reflected in the observed activation energy differences for the $\tau(CH_3)$ mode in TMA and DMA. In addition, DMA cations form N−H⋯N hydrogen bonds with the terminal atoms of the azides. Hydrogen bonds are crucial for stabilizing the crystal structure and defining critical structural phase transition temperatures of these compounds [67]. While the DMA cations are well ordered in the low and high temperature phases in DMACd($N_3$)$_3$, the TMA$^+$ cations, which do not form hydrogen bonds, undergo an abrupt increase in structural disorder when the TMACd($N_3$)$_3$ goes to the high phase. This evidences that hydrogen bonds display a fundamental role in the order-disorder process observed in these compounds. In addition, the activation energy values extracted in this work are comparable to those previously reported for a variety of TMA- and DMA-based compounds. As summarized in **Table S2**, typical activation energies (in meV) are listed for TMA and DMA rotations, TMA Raman-based modes, and CH₃ group rotations. The values obtained in this study (highlighted in blue) fall within the same order of magnitude, supporting the consistency of the observed dynamics with known molecular motion in similar hybrid systems.

**Table 2**. Values of the activation energy calculated from Eq. (10), where the calculated $\Gamma_{EF}$ (Eq. 4) was used in the temperature intervals indicated in the IT phase below the critical temperature ($T_C = 322\ K$) in TMACd($N_3$)$_3$.

| Modes | | EF model | | |
|---|---|---|---|---|
| Assignment | w (cm$^{-1}$) | U (meV) | ln $C$ | Temp. Interval (K) |
| L($N_3^-$) | 221 | 17.43 | 4.02 | 305<T<322 |
| | | 3.18 | 3.48 | 270<T<300 |
| $\tau(CH_3)$ | 275 | 24.67 | 3.40 | 305<T<322 |

| | | 4.48 | 3.24 | 270<T<300 |
|---|---|---|---|---|
| $v_s(NC_4)$ | 751 | 5.89 | 2.08 | 305<T<322 |
| | | 1.07 | 1.89 | 270<T<300 |
| $v_s(N_3)$ | 1360 | 8.79 | 2.22 | 305<T<322 |
| | | 1.55 | 1.95 | 270<T<300 |
| $v_s(CH_3)$ | 2954 | 3.17 | 2.88 | 305<T<322 |
| | | 0.60 | 2.78 | 270<T<300 |
| $v_{as}(CH_3)$ | 3032 | 2.79 | 2.57 | 305<T<322 |
| | | 0.54 | 2.48 | 270<T<300 |

At this point, it is particularly interesting to discuss the activation energy calculated for the librational mode L($N_3^-$) close to $T_C$ (~17.4 meV), which is approximately twice as high as that observed for the symmetrical stretching mode $v_sN_3$ (~8.8 meV). For this, we have observed the internal vibration features of the azide ligands in the low temperature (IT) and high temperature (HT). As shown in figure 4a-b, in the low-temperature phase (ferroelastic), the azide ligands are ordered along the axes $a$ e $c$ (namely Cis – EE azide), and they are disordered along the axis $b$ (namely Trans – EE azide). Since the activation energy calculation depends on the order parameter $Q$, below $T_C$, we analyze the phase transition associated with the changes observed in the Cis – EE azide chains. As shown in Figure 4b, in the high temperature phase (paraelastic) the azide ligand presents high disorder, where the N ions begin to occupy 4 independent Wyckoff positions. This phase can be further characterized by the rotation of the azide ligand along the three axes [68] which are supposed to cause a strong impact in the librational mode L($N_3^-$). It is important to highlight that the cis–EE azide chains, associated with the librational motion of the $N_3$ groups, contribute to positive thermal expansion (PTE) along the $a$ and $c$ axes, which is more pronounced than the negative thermal expansion (NTE) observed along the $b$ axis in the trans–EE azide chains (see Supplementary Figure S3 for a schematic representation of the strain geometry).

The symmetry-breaking strain $e_t = \frac{1}{\sqrt{3}}(2e_3 - e_1 - e_2)$ reflects the anisotropic lattice distortion resulting from the coexisting Cis- and Trans-EE azide chains in the low-temperature ferroelastic phase. Below $T_C$, the elongation along the $a$ and $c$ axes (PST) and the anomalous contraction along $b$ (NTE) give rise to a pronounced $e_t$ (~6%), indicative of both ferroelastic distortion and conformational order. The magnitude of $e_t$ obtained in our study is comparable to the spontaneous strain observed in the ferroelastic formate-based compound [(CH$_2$)$_3$NH$_2$][Mn(HCOO)$_3$] (~5%) [69], and even larger than the shear strains typically below 1% observed in improper ferroelastic perovskites such as (Ca,Sr)TiO₃ [70], and approaches the values as high as 5–10% commonly found in Jahn–Teller-driven perovskites [71]. At the transition temperature $T_C$, $e_t$ drops discontinuously, signaling the collapse of this anisotropic distortion. This behavior is consistent with a first-order transition to a high-temperature, disordered phase in which the distinction between Cis and Trans azide chains is lost, and the average symmetry increases. Therefore, $e_t$ serves as a secondary order parameter that tracks the structural and conformational changes across the ferroelastic transition. Taken together, these findings suggest that the librational mode L(N₃⁻), with its higher activation energy and strong coupling to both strain and disorder, plays a central role in the order–disorder ferroelastic transition of TMACdN₃.

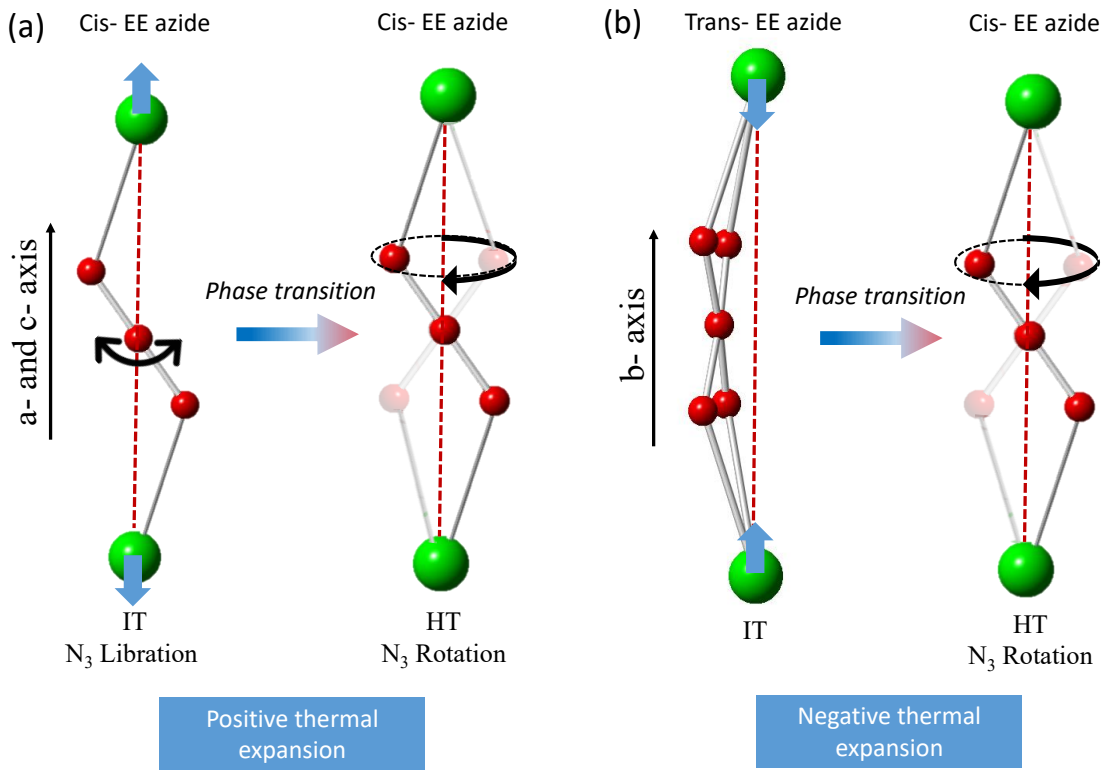

**Figure. 4.** Representation of the evolution of the azide ligands as a function of temperature. a) Schematic representation of the librational mode in the cis-EE azide ligands at intermediate temperature (IT) and the rotational mode at high temperature (HT), associated with large positive thermal expansion (PTE) from IT to HT. b) Schematic representation of the Trans-EE azide ligands at intermediate temperature (IT) and the rotational mode at high temperature (HT), associated with negative thermal expansion (NTE) from IT to HT.

## 4. CONCLUSIONS

Based on the energy fluctuation (EF) model, we investigated the molecular dynamics associated with the ferroelastic phase transition of $TMACd(N_3)_3$. By analyzing selected phonon modes, we calculated the activation energies within the temperature region below $T_C$ in $TMACd(N_3)_3$. Such energies were found to be higher than those obtained for the $DMACd(N_3)_3$. These differences have been mainly attributed to the absence of hydrogen bonds in $TMACd(N_3)_3$, and the larger volume of the TMA spheroidal cation in the $Cd(N_3)_3$ cavity. Our results indicate that hydrogen bonds play a predominant role in order-disorder processes in these systems and in modulating the potential barrier. Importantly, the ferroelastic nature of the transition implies the spontaneous development of strain components coupled to symmetry-lowering distortions. In this context, the $\tau CH_3$ and $L(N_3^-)$ vibrational modes, particularly those associated with the Cis-EE azide chain, are strongly coupled to symmetry-breaking strain, thereby playing a central role in the order–disorder ferroelastic transition of $TMACd(N_3)_3$. Finally, the relaxation times calculated as a function of temperature using EF model are comparable with those recently calculated for $DMACd(N_3)_3$.

**Acknowledgments**

The authors thank the Conselho Nacional de Desenvolvimento Científico e Tecnológico — CNPq (Grants 406322/2022-8) for financial support.